\documentclass[preprint]{imsart}

\arxiv{math.PR/0000000}

\usepackage{amsmath}
\usepackage{amsthm}
\usepackage{amsfonts}
\usepackage{amssymb}
\theoremstyle{plain}
\newtheorem{thm}{Theorem}[section]
\newtheorem{assu}[thm]{Assumption}
\newtheorem{alg}[thm]{Algorithm}

\newtheorem{lemma}[thm]{Lemma}
\newtheorem{cor}[thm]{Corollary}
\theoremstyle{definition}

\theoremstyle{remark}
\newtheorem{remark}[thm]{Remark}

\def\stany{\mathcal{X}}

\def\1c{\mathbb{I}_C(x)}

\begin{document}

\begin{frontmatter}

\title{Rigorous confidence bounds for MCMC under a geometric drift condition\protect\thanksref{T1}}
\runtitle{Rigorous MCMC under Drift Condition}
\thankstext{T1}{Work partially supported by Polish Ministry of Science and Higher Education  Grant No. N N201387234.}

\begin{aug}
\author{\fnms{Krzysztof} \snm{{\L}atuszy\'{n}ski} 
\ead[label=e1]{latuch@gmail.com}}
\and
\author{\fnms{Wojciech} \snm{Niemiro}\corref{}
\ead[label=e3]{wniem@mat.uni.torun.pl}
\ead[label=u1,url]{http://www2.warwick.ac.uk/fac/sci/statistics/staff/research/latuszynski/}}

\runauthor{K. {\L}atuszy\'nski et al.}

\affiliation{University of Warwick and Nicolaus Copernicus University}

\address{K. {\L}atuszy\'nski\\ Department of Statistics\\
University of Warwick\\ CV4 7AL, Coventry, UK\\
\printead{e1}\\
\printead{u1}}

\address{W. Niemiro\\Faculty of Mathematics and Computer Science\\ Nicolaus Copernicus University\\ 
Chopina 12/18, 87-100 Toru\'{n}, Poland\\
\printead{e3}}
\end{aug}

\begin{abstract}
We assume a drift condition towards a small set and bound the mean
square error of estimators obtained by taking averages along a
single trajectory of a Markov chain Monte Carlo algorithm. We use
these bounds to construct fixed-width nonasymptotic confidence intervals. For a
possibly unbounded function $f:\stany \to R,$ let $I=\int_{\stany}
f(x) \pi(x) dx$ be the value of interest and
$\hat{I}_{t,n}=(1/n)\sum_{i=t}^{t+n-1}f(X_i)$ its MCMC estimate. Precisely,
we derive lower bounds for the length of the trajectory $n$ and burn-in
time $t$ which ensure that
$$P(|\hat{I}_{t,n}-I|\leq \varepsilon)\geq 1-\alpha.$$
The bounds depend only and explicitly on drift parameters, on the $V-$norm of $f,$ where $V$ is the drift function and on precision and confidence parameters $\varepsilon, \;\alpha.$ Next we analyse an MCMC estimator based on the
median of multiple shorter runs that allows for sharper bounds for
the required total simulation cost. In
particular the methodology can be applied for computing Bayesian
estimators in practically relevant models. We illustrate our bounds numerically in a simple example.
\end{abstract}

\begin{keyword}[class=AMS]
\kwd[Primary ]{60J10, 65C05} \kwd[; secondary ]{62F15}
\end{keyword}

\begin{keyword}
\kwd{MCMC estimation} \kwd{confidence intervals} \kwd{mean square error} \kwd{Markov
chain} \kwd{convergence rate} \kwd{V-uniform ergodicity}
\kwd{drift condition} \kwd{simulation cost}
\end{keyword}

\end{frontmatter}

\section{Introduction}
An essential part of many problems in Bayesian inference is the
computation of analytically intractable integral
\[ I=\int_{\mathcal{X}} f(x) \pi (x)dx, \] where $f(x)$ is the target function
of interest, $\mathcal{X}$ is often a region in high-dimensio\-nal
space and the probability distribution $\pi$ over $\mathcal{X}$ is
usually known up to a normalizing constant and direct simulation
from $\pi$ is not feasible (see e.g. \cite{RoCa}, \cite{Liu}). A
common approach to this problem is to simulate an ergodic Markov
chain $(X_{n})_{n\geq 0}$, using a transition kernel $P$ with
stationary distribution $\pi$, which ensures that $X_{n} \to \pi$ in distribution. Thus, for
a "large enough" $n_{0}$, $X_{n}$ for $n \geq n_{0}$ is approximately distributed as $\pi$.  Since a simple and powerful
algorithm has been introduced in 1953 by Metropolis et al. in a
very seminal paper \cite{Metrop}, various sampling schemes and
approximation strategies have been developed and analyzed
(\cite{Liu}, \cite{RoCa}) and the method is referred
to as Markov chain Monte Carlo (MCMC).

The standard method is to use average along a single trajectory of the underlying Markov
chain and discard the initial part to reduce bias. In this case
the estimate is of the form
\begin{equation}\label{est_along_one_walk}
\hat{I}_{t,n}=\frac{1}{n}\sum_{i=t}^{t+n-1}f(X_i)\end{equation}
and $t$ is called the burn-in time. Asymptotic validity of  (\ref{est_along_one_walk}) is ensured by a law of large numbers that holds in this setting under very mild assumptions \cite{RobRos survey}. Various results justify the choice of (\ref{est_along_one_walk}). In particular, for reversible chains, Geyer
in \cite{Geyer practical} shows that subsampling is ineffective (in
terms of asymptotic variance) and Chan and Yue in \cite{chan yue -
jrssb - asymptotic} consider 
asymptotic efficiency of (\ref{est_along_one_walk}) in a class of linear estimators (in terms
of mean square error). Asymptotic behaviour of $\hat{I}_{t,n}$ is usually examined via a Central Limit Theorem (CLT) for Markov chains c.f. \cite{Geyer practical, Jones_CLT_surv, RobRos survey}. One constructs asymptotic confidence intervals, based on the CLT and consistent estimators of the asymptotic variance, as described in \cite{Geyer practical, JonesNaranCaffoNeath, HoJo_honest, BeCle2}. Asymptotic behaviour of the mean square error of $\hat{I}_{0,n}$ in the $V-$uniformly ergodic setting has been also studied by Math\'e in \cite{Mathe} using arguments from interpolation theory. 

The goal of this paper is to derive explicit lower bounds for $n$ and $t$
in (\ref{est_along_one_walk}) that ensure the following condition:
 \begin{equation} \label{eps-alpha} P(|\hat{I}_{t,n}-I|\leq \varepsilon)\geq
1-\alpha,\end{equation} where $\varepsilon$ is the precision of
estimation and $1-\alpha,$ the confidence level. We insist on obtaining bounds which depend only on $\varepsilon, \alpha$ and computable characteristics of the transition kernel $P$ and function $f.$  To decrease the total simulation
cost, apart from $\hat{I}_{t,n},$ we also consider a nonlinear
estimator based on the median of multiple shorter runs.

Results of this or related type have been obtained for finite or compact 
state space $\mathcal{X}$ and bounded target function $f$ in \cite{Aldous, Gillman, Rud}. Niemiro and Pokarowski in
\cite{NiePo} give results for relative precision estimation. For
uniformly ergodic chains and bounded function $f,$ Hoeffding type inequalities are
available in \cite{Glynn, MeynKonto, Perron} and can easily lead
to (\ref{eps-alpha}).

Tail inequalities for bounded functionals of Markov chains that are not uniformly ergodic were considered in \cite{Cle}, \cite{Adamczak} and \cite{DoGuMou} using regeneration techniques. Computing explicit bounds from these results may be possible with additional work, but we do not pursue it here.

If the target function $f$ is not bounded and the Markov chain is not uniformly ergodic, rigorous nonasymptotic results about finite sample behaviour of $\hat{I}_{t,n}$ are scarce. Tail inequalities valid in this setup have been established by Bertail and Cl\'emen\c con in \cite{BeCle3} by regenerative approach and using truncation arguments. However, they involve non-explicit constants and can not be directly applied to derive lower bounds on $t$ and $n.$ In \cite{LaMiaNie} a result analogous to (\ref{eps-alpha}) is established for a sequential-regenerative estimator (instead of $\hat{I}_{t,n}$). The approach of \cite{LaMiaNie} requires identification of regeneration times. In many problems of practical interest, especially in high dimension, regeneration schemes are difficult to implement \cite{GilksRobertsSahu, SahuZhigljavsky}.

Our approach is to assume a version of the well known drift
condition towards a small set (Assumption \ref{ass:drift} in
Section \ref{sec:drift}), which is the typical setting when
dealing with integrals of unbounded functions on noncompact sets.
Under this assumption in \mbox{Section \ref{sec:MainResult}} we bound the
mean square error of $\hat{I}_{t,n}.$ Our main Theorem \ref{MSE bound general} exploits the result of Baxendale \cite{Bax}. In Section \ref{sec:eps-alpha} we
study confidence estimation (\ref{eps-alpha}) and obtain explicit
lower bounds on $n$ and $t$ in terms of drift parameters defined
in Assumption \ref{ass:drift}, the $V-$norm of $f,$ where $V$ is the drift function (for definitions see Section \ref{Sec:notation}) and  estimation parameters 
$\varepsilon,$ $\alpha.$ Our bounds
are designed to minimise the total simulation cost $t+n.$ The estimation scheme is then refined via an elementary
exponential inequality for a nonlinear estimator, a median of
multiple shorter runs. In Section \ref{sec: Toy Example} we give
an illustrative toy example. 

The emphasis in our paper is on unbounded $f,$ noncompact $\mathcal{X}$ and nonuniformly ergodic Markov chains, because this is a setting which usually arises when computing Bayesian estimators in many practically relevant models. We note that drift conditions required to apply our approach have been established in particular for the important hierarchical random effects models in \cite{Jones Hobert Gibbs for rand eff mod} and for a more general family of linear models in \cite{Johnson Jones}.

\subsection{Notation and Basic Definitions}\label{Sec:notation}
Throughout this paper, $\pi$ represents the probability measure of
interest, defined on some measurable state space
$(\mathcal{X},\mathcal{F})$ and $f:\mathcal{X}\to R,$ the target
function. Let $(X_n)_{n\geq0}$ be a time homogeneous Markov chain
on $(\mathcal{X},\mathcal{F})$ with transition kernel $P.$ By $\pi_0$ denote its initial
distribution and by $\pi_t$ its distribution at time $t.$ Let
$I=\int_{\mathcal{X}} f(x) \pi (dx)$ be the value of interest and
$\hat{I}_{t,n}=\frac{1}{n}\sum_{i=t}^{t+n-1}f(X_i)$ its MCMC
estimate along one walk.

For a probability measure $\mu$ and a transition kernel $Q$, by
$\mu Q$ we denote a probability measure defined by $\mu
Q(\cdot):=\int_{\mathcal{X}} Q(x,\cdot)\mu (dx).$ In this convention $\pi_t = \pi_0 P^t.$ Furthermore if
$g$ is a real-valued function on $\mathcal{X},$ let
$Qg(x):=\int_{\mathcal{X}} g(y)Q(x,dy)$ and $\mu
g:=\int_{\mathcal{X}} g(x)\mu(dx)$. We will also use $E_{\mu}g$
for $\mu g.$ If $\mu=\delta_x$ we will write $E_x$ instead of $E_\mu.$
For transition kernels $Q_1$ and $Q_2$, $Q_1Q_2$ is also a
transition kernel defined by $Q_1Q_2(x,\cdot):=\int_{\mathcal{X}}
Q_2(y,\cdot)Q_1(x,dy)$.

Let $V:\mathcal{X}\to [1,\infty)$ be a measurable function. For a
measurable function $g:\mathcal{X} \to R$ define its \emph{V-norm}
as
\[ |g|_{V}:=\sup_{x \in
\mathcal{X}}\frac{|g(x)|}{V(x)}. \]
To evaluate the distance between two probability measures
$\mu_{1}$ and $\mu_{2}$ we use the \emph{V-norm distance}, defined as
\[ \|\mu_{1}-\mu_{2}\|_{V} := \sup_{|g| \leq V}
\left|\mu_1g-\mu_2g
\right|.\]
Note that for $V\equiv 1$ the $V-$norm distance $||\cdot||_{V}$
amounts to the well known total variation distance, precisely
$\|\mu_{1}-\mu_{2}\|_{V}=2||\mu_{1}-\mu_{2}||_{\textrm{tv}}:=2\sup_{A\in
\mathcal{F}} |\mu_{1}(A)-\mu_{2}(A)|.$

 Finally for two transition kernels $Q_1$ and $Q_2$ the \emph{V-norm
 distance} between $Q_1$ and $Q_2$ is defined by
 \[|||Q_1-Q_2|||_{V}:=\big|\|Q_1(x,\cdot)-Q_2(x,\cdot)\|_{V}\big|_{V}=
 \sup_{x\in
 \mathcal{X}}\frac{\|Q_1(x,\cdot)-Q_2(x,\cdot)\|_{V}}{V(x)}.
 \]
 For a probability distribution $\mu,$ define a transition
 kernel $\mu(x,\cdot):=\mu(\cdot),$ to allow for writing
 $|||Q-\mu|||_{V}$ and $|||\mu_1-\mu_2|||_{V}.$
 Define also \[B_V:=\{f:f:\mathcal{X}\to R, |f|_{V}<\infty\}.\]
 Now if $|||Q_1-Q_2|||_{V}<\infty,$ then $Q_1-Q_2$ is a bounded operator
 from $B_V$ to itself, and $|||Q_1-Q_2|||_{V}$ is its operator norm.
 See \cite{MeynTweedie} for details.

In the sequel we will work with geometrically ergodic Markov chains. A Markov chain is said to be \emph{geometrically ergodic} if 
\[ \|\delta_x P^n-\pi\|_{\textrm{tv}} \leq M(x)\tilde{\gamma}^n, \quad \textrm{for} \quad \pi-\textrm{a.e.}\; x, \quad \textrm{and for some} \quad \tilde{\gamma}<1.\]
In particular, if $M(x) \leq M$ then the chain is said to be uniformly ergodic. Geometric ergodicity is equivalent to existence of a drift function $V$ towards a small set (see \cite{RobRos survey} and c.f. Assumption \ref{ass:drift}) and consequently also to \emph{$V-$uniform ergodicity} which is defined by the following condition.
\[ \|\delta_x P^n-\pi\|_{V} \leq MV(x)\gamma^n \quad \textrm{or equivalently} \quad |||P^n-\pi |||_{V} \leq M\gamma^n,\]
for some $M < \infty$ and some $\gamma < 1.$ 
\section{A Drift Condition and Preliminary Lemmas}
\label{sec:drift}
We analyze the MCMC estimation under the
following assumption of a drift condition towards a small set, c.f. \cite{Bax}.
\begin{assu} \label{ass:drift}~
\begin{itemize}
\item[(A.1)]{Small set.} There exist $C\subseteq \mathcal{X},$ $\tilde{\beta}>0$ and a
probability measure $\nu$ on $\mathcal{X},$ such that for all
$x\in C$ and $A \subseteq \mathcal{X}$ $$P(x,A)\geq \tilde{\beta}
\nu (A).$$
\item[(A.2)]{Drift.} There exist a function $V:\mathcal{X} \to
[1,\infty)$ and constants $\lambda<1$ and $K<\infty$ satisfying
$$PV(x) \leq \left\{\begin{array}{lcc}
\lambda V(x), & \text{if} & x\notin C, \\ K, & \text{if} & x \in
C.
\end{array}
\right.$$
\item[(A.3)]{Strong Aperiodicity.} There exists $\beta>0$ such that
$\tilde{\beta} \nu(C) \geq \beta.$
\end{itemize}
\end{assu}

In the sequel we refer to $\tilde{\beta}, V, \lambda, K, \beta$ as
drift parameters.

This type of drift condition is often assumed and widely discussed
in Markov chains literature since it implies geometric ergodicity and a CLT for a convenient class of target functions, see \cite{MeynTweedie} for details and definitions. Computable
bounds for geometric ergodicity parameters under drift conditions allow to control the burn-in time $t$ and the bias of MCMC estimators in practically relevant models. Substantial effort has been devoted
to establishing such bounds, c.f. the survey paper by Roberts and
Rosenthal \cite{RobRos survey} and references therein.  Particular references include e.g.  Rosenthal
\cite{Rosenthal drift} or Roberts and Tweedie \cite{Roberts
Tweedie} for bounds on the total variation distance. Since we deal with unbounded functions, in the sequel we
make use of the $V-$uniform ergodicity convergence bounds obtained
by Baxendale in \cite{Bax} (c.f. Douc \textit{at al.} \cite{DMR} and Fort \cite{Fort}). In the drift condition setting and using explicit convergence bounds, our goal is to control not only the burn-in time $t,$ but also the length of simulation $n.$
\begin{thm}[\cite{MeynTweedie},\cite{Bax}] \label{thm:bax}
Under Assumption \ref{ass:drift} $(X)_{n\geq 0}$ has a unique
stationary distribution $\pi$ and $\pi V < \infty$ (\cite{MeynTweedie}). Moreover (Theorem 1.1 of \cite{Bax}),
there exists $\rho < 1$ depending only and explicitly on
$\tilde{\beta}, \beta, \lambda$ and $K$ such that whenever $\rho <
\gamma < 1$ there exists $M < \infty$ depending only and
explicitly on $\gamma, \tilde{\beta}, \beta, \lambda$ and $K$ such
that for all $n\geq 0$
 \begin{equation}\label{eqn: bax}
 |||P^n-\pi |||_{V} \leq M\gamma^n.\end{equation}
\end{thm}

Formulas for $\rho = \rho(\tilde{\beta}, \lambda, K, \beta)$
and $M = M(\gamma,\tilde{\beta}, \lambda, K, \beta)$ established in \cite{Bax} are given in Appendix \ref{sec:Bax form} and are used in
Section \ref{sec: Toy Example}. To our
knowledge the above-mentioned theorem gives the best available
explicit constants. However this is a topic of ongoing research (c.f. \cite{Bednorz bounds}). We note that improving ergodicity constants in Theorem \ref{thm:bax} will automatically result in tightening bounds established in our paper.

\begin{cor} \label{cor: burn-in} Under
Assumption \ref{ass:drift}
\[\|\pi_0 P^n-\pi \|_{V} \leq
 \min\{\pi_0 V, \|\pi_0 - \pi \|_{V}\} M\gamma^n,\]
where $M$ and $\gamma$ are such as in Theorem \ref{thm:bax}.
\end{cor}

\begin{proof} From Theorem \ref{thm:bax} we have
$\|P^n(x,\cdot)-\pi(\cdot)\|_{V} \leq M\gamma^n V(x),$ which
yields \begin{eqnarray} \pi_0 V M \gamma^n & \geq &
\int_{\mathcal{X}} \|P^n(x,\cdot)-\pi(\cdot)\|_{V} \pi_0(dx) \geq
\sup_{|g|\leq V} \int_{\mathcal{X}}|P^n(x,\cdot)g-\pi g|\pi_0(dx)
\nonumber \\ & \geq & \sup_{|g|\leq V} |\pi_0 P^ng-\pi g|= \|\pi_0
P^n-\pi \|_{V}.\nonumber \end{eqnarray}
 Now let $b_V=\inf_{x\in \mathcal{X}}V(x)$ and let $\mu_1,$ $\mu_2$ be measures. Clearly $\|\mu_1(x,\cdot) - \mu_2(x, \cdot)\|_V$ is constant in $x$ and
therefore \[ ||| \mu_1 - \mu_2 |||_V = \sup_x \frac{\|\mu_1(x,\cdot) - \mu_2(x,\cdot)\|_V}{V(x)} = \frac{\|\mu_1 - \mu_2\|_V}{b_V}. \]
Since $|||\cdot|||_{V}$ is an operator norm and
$\pi$ is invariant for $P$, we have
\begin{eqnarray} \|\pi_0 P^n-\pi \|_{V} & = & b_V|||\pi_0 P^n-\pi
|||_{V} = b_V|||(\pi_0 - \pi)(P^n-\pi) |||_{V} \nonumber \\ & \leq
& b_V|||\pi_0 - \pi|||_{V}|||P^n-\pi |||_{V} = \|\pi_0 - \pi
\|_{V}|||P^n-\pi |||_{V}. \nonumber \\ & \leq & \|\pi_0-\pi\|_V
M\gamma^n. \nonumber \end{eqnarray}
\end{proof}

Next we focus on the following simple but useful observation.
\begin{lemma} \label{lem:hoelder} If for a Markov chain $(X_n)_{n \geq 0}$ on
$\mathcal{X}$ with transition kernel $P$ Assumption
\ref{ass:drift} holds with parameters $\tilde{\beta}, V, \lambda,
K, \beta,$ it holds also with $\tilde{\beta}_r:=\tilde{\beta},$
$V_r:=V^{1/r},$ $\lambda_r:=\lambda^{1/r},$ $K_r:=K^{1/r},$
$\beta_r:=\beta$ for every $r>1.$
\end{lemma}
\begin{proof}
It is enough to check (A.2). For $x \notin C$ by Jensen inequality
we have
\[ \lambda V(x)\geq 
\int_{\mathcal{X}} V(y)P(x,dy) \geq \left(\int_{\mathcal{X}}
V(y)^{1/r}P(x,dy)\right)^{r}
\]
and hence $PV_r(x)\leq \lambda^{1/r}V_r(x),$ as claimed. Similarly
for $x\in C$ we obtain $PV_r(x)\leq K^{1/r}.$
\end{proof}

Lemma \ref{lem:hoelder} together with Theorem \ref{thm:bax} yield
the following corollary.
\begin{cor} \label{cor:hoelder} Under Assumption \ref{ass:drift}
we have
\[|||P^n-\pi |||_{V^{1/r}} \leq M_r \gamma_r^n,\]
where $M_r$ and $\gamma_r$ are constants defined as in Theorem
\ref{thm:bax} resulting from drift parameters defined in Lemma
\ref{lem:hoelder}.
\end{cor}

Integrating the drift condition with respect to $\pi$ yields the
following bound on $\pi V.$
\begin{lemma} \label{lemma: pi V bound}
Under Assumption  \ref{ass:drift}
\[ \pi V \leq \pi(C) \frac{K-\lambda}{1-\lambda} \leq \frac{K-\lambda}{1-\lambda}. \]
\end{lemma}

Let $f_c = f - \pi f.$ The next lemma provides a bound on
$||f_c|^p|_V$ in terms of $||f|^p|_V$ without additional effort.
\def\cp{||f|^p|_V}
\begin{lemma} \label{lemma: f_c^p V-norm}
Under Assumption  \ref{ass:drift}
\[
||f_c|^p|_V^{2/p} \leq \Big(\cp^{1/p} + \frac{\pi
(C)}{b_V^{1/p}} K_{p, \lambda} \Big)^2
 \leq \big(\cp^{1/p} + K_{p, \lambda} \big)^2,
\]
 where $b_V = \inf_{x \in \stany}
V(x)$ and $K_{p,\lambda} =
\frac{K^{1/p}-\lambda^{1/p}}{1-\lambda^{1/p}}.$
\end{lemma}

\begin{proof} Note that $\pi
V^{1/p} \leq \pi (C) K_{p,\lambda} \leq K_{p, \lambda}$ by Lemma
\ref{lemma: pi V bound} and proceed:
\begin{eqnarray}
||f_c|^p|_V & = & \sup_{x \in \stany} \frac{|f(x) - \pi
f|^p}{V(x)} \leq \sup_{x \in \stany} \frac{\Big(\cp^{1/p}
V^{1/p}(x) + \pi |f|\Big)^p}{V(x)} \nonumber \\ \nonumber
 & \leq & \sup_{x \in \stany} \frac{\Big(\cp^{1/p}
V^{1/p}(x) + \pi (C) K_{p,\lambda}\Big)^p}{V(x)}
 \leq \cp \bigg(1+
 \frac{\pi (C) K_{p, \lambda}}{b_V^{1/p} \cp^{1/p}
 }\bigg)^p.
\end{eqnarray}
\end{proof}

\section{MSE Bounds} \label{sec:MainResult}

By $MSE(\hat{I}_{t,n})$ we denote the mean square error of
$\hat{I}_{t,n},$ i.e.
$$MSE(\hat{I}_{t,n})=E_{\pi_0}[\hat{I}_{t,n}-I]^2.$$
Nonasymptotic bounds on
$MSE(\hat{I}_{t,n})$ are essential to establish
confidence estimation (\ref{eps-alpha}) and
are also of independent interest. The main result of this section is the following
\begin{thm}[MSE Bounds]\label{MSE bound general}
Assume the Drift Condition \ref{ass:drift} holds and $X_0\sim
\pi_0.$ Then for every measurable function $f:\mathcal{X}\to R,$
every $p\geq 2$ and every $r\in [\frac{p}{p-1},p]$
\begin{equation}\label{MSE bound general bound}
MSE(\hat{I}_{0,n}) \leq
\frac{||f_c|^p|_V^{2/p}}{n}\left(1+\frac{2M_r
\gamma_r}{1-\gamma_r}\right)\left(\pi V + \frac{M\min\{\pi_0
V,\|\pi_0-\pi\|_{V}\}}{n(1-\gamma)}\right),\end{equation}
 where $f_c=f-\pi f$ and constants $M, \gamma, M_r, \gamma_r$ depend
 only and explicitly on
 $\tilde{\beta}, \beta, \lambda$ and $K$ from Assumption \ref{ass:drift} as
 in Theorem \ref{thm:bax} and Corollary \ref{lem:hoelder}.
\end{thm}
We emphasise the most important special case for $p=r=2$ as a corollary.
\begin{cor}\label{MSE bound typical}
In the setting of Theorem \ref{MSE bound general}, we have in
particular
\begin{equation} \label{MSE bound typical bound} MSE(\hat{I}_{0,n})\leq
\frac{|f_c^2|_V}{n}\left(1+\frac{2M_2
\gamma_2}{1-\gamma_2}\right)\left(\pi V + \frac{M\min\{\pi_0
V,\|\pi_0-\pi\|_{V}\}}{n(1-\gamma)}\right).\end{equation}
\end{cor}
\begin{remark} The formulation of the foregoing Theorem \ref{MSE bound general}
is motivated by a trade-off between small $V$ and small $\lambda$
in Assumption \ref{ass:drift}. It should be intuitively clear that
establishing the drift condition for a quickly increasing $V$
should result in smaller $\lambda$ at the cost of bigger $\pi V.$
So it may be reasonable to look for a valid drift condition with
$V\geq C||f_c|^p|$ for some $p>2$ instead of the natural choice of
$p=2.$ Lemma \ref{lem:hoelder} should strengthen this intuition. \end{remark}
\begin{remark}
For evaluating $\min\{\pi_0
V,\|\pi_0-\pi\|_{V}\}$ one will often use the obvious bound $\min\{\pi_0
V,\|\pi_0-\pi\|_{V}\}\leq \pi_0 V,$ because $\pi_0 V$
depends on $\pi_0$ which is users choice, e.g. a
deterministic point. Also, in some cases a fairly small bound for
$\pi V$ should be possible to obtain by direct calculations, e.g. if $\pi$ is exponentially concentrated and
$V$ is a polynomial of degree 2. However, in absence of a better bound for $\pi
V,$ Lemma \ref{lemma: pi V bound} is at hand. Similarly Lemma
\ref{lemma: f_c^p V-norm} bounds the unknown value
$||f_c|^p|_V^{2/p}$ in terms of $||f|^p|_V.$ Note that in
applications both $f$ and $V$ have explicit formulas known to the
user and $||f|^p|_V$ can be evaluated directly or easily bounded.\end{remark}
\begin{remark} Let $\sigma^2_{\textrm{as}}(f)$ denote the asymptotic variance from the CLT for Markov chains (see e.g. \cite{RobRos survey, BeLaLa}). Since in the drift condition setting \[\frac{nMSE(\hat{I}_{0,n})}{\sigma^2_{\textrm{as}}(f)} \to 1 \quad \textrm{as} \quad n \to \infty, \] we see that the bounds in Theorem \ref{MSE bound general} and Corollary \ref{MSE bound typical} have the correct asymptotic dependence on $n$ and are easy to interpret. In particular $\pi V|f_c^2|_V$ in Corollary \ref{MSE bound typical} should
be close to $Var_{\pi}f$ for an appropriate choice of $V,$
the term $2M_2\gamma_2/(1-\gamma_2)$ corresponds to the
autocorrelation of the chain and $M\min\{\pi_0
V,\|\pi_0-\pi\|_{V}\}/n(1-\gamma)$ is the price for
nonstationarity of the initial distribution. In fact Theorem \ref{MSE bound
general} with $\pi_0 = \pi$ yields the following bound on the asymptotic variance \[ \sigma^2_{\textrm{as}}(f)=\lim_{n\to \infty}nE_{\pi}[\hat{I}_{0,n}-I]^2 \leq \pi
V||f_c|^p|_V^{2/p}\left(1+\frac{2M_r\gamma_r}{1-\gamma_r}\right).\]
\end{remark}
\begin{proof}[Proof of Theorem \ref{MSE bound general}]
Note that $|f|_{V^{1/r}}^r=||f|^r|_V.$ Without loss of generality
consider $f_c$ instead of $f$ and assume $||f_c|^p|_V=1.$ In this
setting
 $|f_c^2|_V\leq 1,$ $Var_{\pi}f_c=\pi f_c^2 \leq \pi V,$
$MSE(\hat{I}_{0,n})=E_{\pi_0}(\hat{I}_{0,n})^2,$ and also for
every $r\in [\frac{p}{p-1},p],$
$$|f_c|_{V^{1/r}}\leq ||f_c|^{p/r}|_{V^{1/r}}=1\quad
\textrm{and}\quad |f_c|_{V^{1-1/r}}\leq
||f_c|^{p-p/r}|_{V^{1-1/r}}=1.$$
 Obviously
\begin{eqnarray} \label{eqn:MSE deco} nMSE(\hat{I}_{0,n})&=&
\frac{1}{n}\sum_{i=0}^{n-1}E_{\pi_0}f_c(X_i)^2 +
\frac{2}{n}\sum_{i=0}^{n-2}\sum_{j=i+1}^{n-1}
E_{\pi_0}f_c(X_i)f_c(X_j).
\end{eqnarray}
We start with a bound for the first term of the right hand side of
(\ref{eqn:MSE deco}). Since $f_c^2(x)\leq V(x),$ we use Corollary
\ref{cor: burn-in} for $f_c^2.$ Let $C=\min\{\pi_0
V,\|\pi_0-\pi\|_{V}\}$ and proceed
\begin{equation} \label{eqn:MSE first}
\frac{1}{n}\sum_{i=0}^{n-1}E_{\pi_0}f_c(X_i)^2
=\frac{1}{n}\sum_{i=0}^{n-1}\pi_0 P^i f_c^2 \leq \pi
f_c^2+\frac{1}{n}\sum_{i=0}^{n-1}CM\gamma^{i}
 \leq  \pi V + \frac{CM}{n(1-\gamma)}.
\end{equation}
To bound the second term of the right hand side of (\ref{eqn:MSE
deco}) note that $|f_c|\leq V^{1/r}$ and use Corollary
\ref{cor:hoelder}.
\begin{eqnarray}
\frac{2}{n}\sum_{i=0}^{n-2}\sum_{j=i+1}^{n-1}
E_{\pi_0}f_c(X_i)f_c(X_j) & = &
\frac{2}{n}\sum_{i=0}^{n-2}\sum_{j=i+1}^{n-1} \pi_0 \left(P^i
\left(f_cP^{j-i}f_c\right)\right) \nonumber \\
& \leq & \frac{2}{n}\sum_{i=0}^{n-2}\sum_{j=i+1}^{n-1} \pi_0
\left(P^i
\left(|f_c||P^{j-i}f_c|\right)\right) \nonumber \\
&\leq& \frac{2M_r}{n}\sum_{i=0}^{n-2}\sum_{j=i+1}^{\infty}
\gamma_r^{j-i} \pi_0 \left(P^i \left(|f_c|V^{1/r}\right)\right)
\nonumber \\
&\leq& \frac{2M_r \gamma_r}{n(1-\gamma_r)}\sum_{i=0}^{n-2} \pi_0
\left(P^i \left(|f_c|V^{1/r}\right)\right)=\spadesuit \nonumber
\end{eqnarray}
Since $|f_c|\leq V^{1/r}$ and $|f_c|\leq V^{1-1/r},$ also
$|f_cV^{1/r}|\leq V$ and we use Corollary \ref{cor: burn-in} for
$|f_c|V^{1/r}.$
\begin{eqnarray} \label{eqn:MSE second}
\spadesuit &\leq& \frac{2M_r
\gamma_r}{n(1-\gamma_r)}\sum_{i=0}^{n-2} \left(\pi
\left(|f_c|V^{1/r}\right)+CM\gamma^i \right) \leq \frac{2M_r
\gamma_r}{1-\gamma_r}\left(\pi V + \frac{CM
}{n(1-\gamma)}\right).\qquad
\end{eqnarray}

Combine (\ref{eqn:MSE first}) and (\ref{eqn:MSE second}) to obtain
\begin{eqnarray} MSE(\hat{I}_{0,n}) & \leq &
\frac{||f_c|^p|_V^{2/p}}{n}\left(1+\frac{2M_r
\gamma_r}{1-\gamma_r}\right)\left(\pi V + \frac{CM
}{n(1-\gamma)}\right). \nonumber
\end{eqnarray}
\end{proof}

Theorem \ref{MSE bound general} is explicitly stated for
$\hat{I}_{0,n},$ but the structure of the bound is flexible enough
to cover most typical settings as indicated below.

\begin{cor} \label{cor:MSE bounds various}
In the setting of Theorem \ref{MSE bound general},
\begin{eqnarray}\label{MSE bound perfect}
MSE(\hat{I}_{0,n})  &\leq & \frac{\pi
V||f_c|^p|_V^{2/p}}{n}\left(1+\frac{2M_r
\gamma_r}{1-\gamma_r}\right), \quad \textrm{if} \quad
\pi_0=\pi,\\
\label{MSE bound deterministic start} MSE(\hat{I}_{0,n}) &\leq&
\frac{||f_c|^p|_V^{2/p}}{n}\left(1+\frac{2M_r
\gamma_r}{1-\gamma_r}\right)\left(\pi V +
\frac{MV(x)}{n(1-\gamma)}\right), \quad \textrm{if} \quad
\pi_0=\delta_x, \\
\label{MSE bound burn-in} MSE(\hat{I}_{t,n})  &\leq&
\frac{||f_c|^p|_V^{2/p}}{n}\left(1+\frac{2M_r
\gamma_r}{1-\gamma_r}\right)\left(\pi V + \frac{M^2\gamma^t
V(x)}{n(1-\gamma)}\right), \quad \textrm{if} \quad \pi_0=\delta_x. \qquad
 \end{eqnarray}
\end{cor}

\begin{proof}
Only (\ref{MSE bound burn-in}) needs a proof. Note that $X_t \sim
\delta_x P^t.$ Now use Theorem \ref{thm:bax} to see that
$\|\delta_x P^t - \pi \|_{V} \leq M\gamma^tV(x),$ and apply
Theorem \ref{MSE bound general} with $\pi_0 = \delta_x P^t.$
\end{proof}
Bound (\ref{MSE bound perfect}) corresponds to the situation when
a perfect sampler is available and used instead of burn-in. For deterministic start without
burn-in and with burn-in, (\ref{MSE bound deterministic start}) and
(\ref{MSE bound burn-in}) should be applied respectively.

\section{Confidence Estimation}\label{sec:eps-alpha}

Confidence estimation is an easy corollary of $MSE$
bounds by the Chebyshev inequality.

\begin{thm}[Confidence Estimation]\label{thm:eps-alpha}
Under Assumption \ref{ass:drift}, let
 \begin{eqnarray} \label{def: b}
 b &=& \frac{\pi
V||f_c|^p|_V^{2/p}}{\varepsilon^2 \alpha}\left(1+\frac{2M_r
\gamma_r}{1-\gamma_r}\right),
 \\  \label{def: c} c &=& \frac{M\min\{\pi_0
V,\|\pi_0-\pi\|_{V}\}||f_c|^p|_V^{2/p}}{\varepsilon^2
\alpha(1-\gamma)}\left(1+\frac{2M_r\gamma_r}{1-\gamma_r}\right),\\
 \label{def: c(t)}
c(t) &=& \frac{M^2\gamma^t V(x) ||f_c|^p|_V^{2/p}}{\varepsilon^2
\alpha(1-\gamma)}\left(1+\frac{2M_r\gamma_r}{1-\gamma_r}\right),\\
 \label{def: n(t)}
 n(t)&=&\frac{b+\sqrt{b^2+4c(t)}}{2}, \\
\label{def: wave c} \tilde{c} & = & \frac{M^2V(x)
  ||f_c|^p|_V^{2/p}}{\varepsilon^2
\alpha(1-\gamma)}\left(1+\frac{2M_r\gamma_r}{1-\gamma_r}\right).
 \end{eqnarray}
 Then 
\begin{eqnarray}\label{eps-alpha no burn-in}
 P(|\hat{I}_{0,n}-I| \leq \varepsilon) \geq  1-\alpha,&
\textrm{if} &
  X_0 \sim \pi_0, \quad n \geq \frac{b+\sqrt{b^2+4c}}{2}.\\
\label{eps-alpha with burn-in}
 P(|\hat{I}_{t,n}-I| \leq  \varepsilon) \geq 1-\alpha, &
  \textrm{if}
&  \left\{ \begin{array}{l}  X_0 \sim \delta_x, \\
 t \geq \max\left\{0, 
 \log_{\gamma}\left(
 \frac{2+\sqrt{4+b^2\ln^2\gamma}}{\tilde{c}\ln^2\gamma}\right)\right\},
 \quad\\
  n \geq n(t).
  \end{array} \right.
 \end{eqnarray}
\end{thm}

\begin{remark}[Leading term] \label{rem_leading_term} The above bounds in (\ref{eps-alpha with burn-in}) give the
 minimal length of the trajectory $(t+n)$ resulting from
 (\ref{MSE bound burn-in}). The leading term of the bound on $n$ is $$b = \frac{\pi V |f_c^2|_V}{\varepsilon^2 \alpha}(1+ \frac{2M_2
\gamma_2}{1-\gamma_2})$$ (where we took $p=r=2$ for simplicity). Quantity
$\pi V |f_c^2|_V$ should be of the same order as $Var_{\pi}f,$ thus
a term of this form is inevitable in any bound on $n.$ Next, $\varepsilon^{-2}$ which results from
Chebyshev's inequality, is typical and inevitable, too. The factor $\alpha^{-1}$ will be reduced later in this section to $\log(\alpha^{-1})$
for small $\alpha$ by Lemma \ref{lem: median} and Algorithm
\ref{Alg: median}. The term $1+ \frac{2M_2 \gamma_2}{1-\gamma_2}$ which roughly speaking bounds the autocorrelation of the chain, is the
bottleneck of the approach. Here good bounds on $\gamma$ and the
somewhat disregarded in literature $M = M(\gamma)$ are equally important.
Improvements in Baxendale-type convergence bounds may lead to
dramatic improvement of the bounds on the total simulation cost
(e.g. by applying the preliminary results of \cite{Bednorz
bounds}).\end{remark}

\begin{remark} \label{rem: burn-in} The formulation of Theorem
\ref{thm:eps-alpha} indicates how the issue of
a sufficient burn-in should be understood. The common approach is to describe $t$ as \textit{time to stationarity} and to require that $t^*=t(x,\tilde{\varepsilon})$ should be such that
$\rho (\pi, \delta_xP^{t^*})\leq \tilde{\varepsilon}$ (where
$\rho(\cdot, \cdot)$ is a distance function for probability
measures, e.g. total variation distance, or $V-$norm distance).
This approach seems not appropriate for such a natural goal as
fixed precision of estimation at fixed confidence level. The optimal burn-in time can
be much smaller then $t^*$ and in particular cases it can be $0.$
Also we would like to emphasise that in the typical drift
condition setting, i.e. if $\mathcal{X}$ is not compact and the
target function $f$ is not bounded,  $||\pi_t-\pi||_{\rm{tv}} \to 0$ does
not even imply $\pi_tf \to \pi f.$ Therefore a $V-$norm with $|f|_V  < \infty$ should be used as
a measure of convergence.
\end{remark}

\begin{proof}[Proof of Theorem \ref{thm:eps-alpha}]
From the Chebyshev's inequality we get
 \begin{eqnarray} \label{Czybyszew} P(|\hat{I}_{t,n}-I|\leq
\varepsilon) &=& 1- P(|\hat{I}_{t,n}-I|\geq \varepsilon) \nonumber \\
&\geq &1-\frac{MSE(\hat{I}_{t,n})}{\varepsilon^2} \geq  1-\alpha
\quad \textrm{if}\quad MSE(\hat{I}_{t,n})\leq \varepsilon^2
\alpha. \qquad
\end{eqnarray}
To prove (\ref{eps-alpha no burn-in}) set $C=\min\{\pi_0
V,\|\pi_0-\pi\|_{V}\},$ and combine (\ref{Czybyszew}) with
(\ref{MSE bound general bound}) to get $$n^2-n \frac{\pi
V||f_c|^p|_V^{2/p}}{\varepsilon^2
\alpha}\left(1+\frac{2M_r\gamma_r}{1-\gamma_r}\right) -\frac{M
C||f_c|^p|_V^{2/p}}{\varepsilon^2
\alpha(1-\gamma)}\left(1+\frac{2M_r\gamma_r}{1-\gamma_r}\right)\geq
0,$$ and hence $n  \geq \frac{b+\sqrt{b^2+4c}}{2},$ where $b$ and
$c$ are defined by (\ref{def: b}) and (\ref{def: c}) respectively.
The only difference in (\ref{eps-alpha with burn-in}) is that now
we have $c(t)$ defined by (\ref{def: c(t)}) instead of $c.$ It is
easy to check that the best bound on $t$ and $n$ (i.e. which
minimizes $t+n$) is such that
$$n\geq n(t) \qquad \textrm{and} \qquad t \geq \max\left\{0,
\min\{t\in N: n'(t)\geq -1\}\right\},$$
where $n(t)$ is defined by (\ref{def: n(t)}) and $n'(t) = \frac{d}{dt}n(t).$
 Standard calculations show that
 $$\min\{t\in N: n'(t)\geq -1\} =
 \min\{t\in N: (\gamma^t)^2\tilde{c}^2\ln^2\gamma -
 \gamma^t 4 \tilde{c} - b^2 \leq 0\},$$
 where $\tilde{c}$ is defined by (\ref{def: wave c}). Hence we obtain
 $$t \geq \max\left\{0,(\ln\gamma)^{-1}\ln\left(
 \frac{2+\sqrt{4+b^2\ln^2\gamma}}{\tilde{c}\ln^2\gamma}\right)\right\}
 \qquad \textrm{and} \qquad n\geq n(t).$$
 This completes the proof.
\end{proof}

Next we consider an alternative nonlinear estimation scheme, the so called ''median trick" (introduced in \cite{JeVaViz86} in the computational complexity context and further developed in \cite{NiePo}) that allows for
sharper bounds for the total simulation cost needed to obtain
confidence estimation for small $\alpha.$ The following simple lemma is taken from a more general setting of Section 2 in \cite{NiePo}.

\begin{lemma} \label{lem: median}
Let $m \in N$ be an odd number and let $\hat{I}_1,\dots,
\hat{I}_m$ be independent random variables, such that
$P(|\hat{I}_k - I| \leq \varepsilon) \geq 1-a > 1/2,$ for
$k=1,\dots, m.$ Define $\hat{I}:=
\textup{med}\{\hat{I}_1,\dots,\hat{I}_m\}.$ Then
\begin{equation} \label{eqn: median}
P(|\hat{I}-I| \leq \varepsilon) \geq 1 - \alpha, \quad \textrm{if}
\quad m \geq \frac{2\ln (2\alpha)}{\ln [4a(1-a)]}.
\end{equation} \end{lemma}


Hence confidence estimation with parametesrs $\varepsilon, \alpha$ can be obtained by the
following Algorithm \ref{Alg: median}.

\begin{alg}[MA: median of averages] \label{Alg: median} ~
 \begin{enumerate}
\item Simulate $m$ independent runs of length $t+n$ of the
underlying Markov chain, $$X_{0}^{(k)}, \dots, X_{t+n-1}^{(k)},
\quad k=1,\dots, m.$$
\item Calculate $m$ estimates of $I,$ each based on a single run,
$$\hat{I}_{k}=\hat{I}_{t,n}^{(k)}=\frac{1}{n}\sum_{i=t}^{t+n-1}
f(X_{i}^{(k)}), \quad k=1,\dots,m.$$
\item For the final estimate take
$$\hat{I}=\textup{med}\{\hat{I}_1,\dots,\hat{I}_m\}.$$
\end{enumerate}
\end{alg}
Theorem
\ref{thm:eps-alpha} should be used to find $t$ and $n$ that
guarantee confidence estimation with parameters $\varepsilon, a$ and $m$ results from
Lemma \ref{lem: median}. The total cost of Algorithm \ref{Alg: median} amounts to
$m(t+n)$
 and depends on $a$ (in addition to previous
parameters). The optimal $a$ can be found numerically, however $a=0.11969$ is an acceptable arbitrary choice
(cf. \cite{NiePo}).

\section{A Toy Example - Contracting Normals} \label{sec: Toy Example}

To illustrate the results of previous sections we analyze the
\textit{contracting normals} example studied by Baxendale in
\cite{Bax} (see also \cite{Roberts Tweedie}, \cite{Roberts
Rosenthal shift} and \cite{Rosenthal gibbs}), where Markov chains
with transition probabilities $P(x,\cdot) = N(\theta x, 1-
\theta^2)$ for some parameter $\theta \in (-1,1)$ are considered.

Similarly as in \cite{Bax} we take a drift function $V(x) = 1+
x^2$ and a small set $C = [-d,d]$ with $d > 1,$ which allows for
$\lambda = \theta^2 + \frac{2(1-\theta^2)}{1+d^2} < 1$ and $K = 2
+ \theta^2(d^2-1).$ We also use the same minorization condition
with $\nu$ concentrated on $C,$ such that $\tilde{\beta} \nu (dy)
= \min_{x \in C} (2\pi (1-\theta^2))^{-1/2} \exp (-\frac{(\theta x
- y)^2}{2(1-\theta^2)})dy.$ This yields $\tilde{\beta} =
2[\Phi(\frac{(1+|\theta|)d}{\sqrt{1-\theta^2}}) -
\Phi(\frac{|\theta|d}{\sqrt{1-\theta^2}})],$ where $\Phi$ denotes
the standard normal cumulative distribution function.

 Baxendale in \cite{Bax} indicated that the chain is
reversible with respect to its invariant distribution $\pi =
N(0,1)$ for all $\theta \in (-1, 1)$ and it is reversible and
positive for $\theta > 0.$ Moreover, in Lemma \ref{lemma: contracting normals convergence} we
observe a relationship between marginal distributions of the chain
with positive and negative values of $\theta.$ By $\mathcal{L}(X_n
| X_0, \theta)$ denote the distribution of $X_n$ given the
starting point $X_0$ and the parameter value $\theta.$
\begin{lemma} \label{lemma: contracting normals convergence}
\begin{equation}
\mathcal{L}(X_n|X_0, \theta) = \mathcal{L}(X_n|(-1)^n X_0, -
\theta).
\end{equation} \end{lemma}
\begin{proof}
Let $Z_1, Z_2, \dots $ be an iid $N(0,1)$ sequence, then
\begin{eqnarray}
\mathcal{L}(X_n|X_0, \theta) & = & \mathcal{L}\Big(\theta^n X_0 +
\sum_{k=1}^n \theta^{n-k} \sqrt{1-\theta^2} Z_k \Big)
\nonumber \\
& = & \mathcal{L}\Big((-\theta)^n (-1)^n X_0 + \sum_{k=1}^n
(-\theta)^{n-k} \sqrt{1-\theta^2} Z_k \Big)
\nonumber \\
 &=& \mathcal{L}(X_n|(-1)^n
X_0, - \theta), \nonumber
\end{eqnarray}
and we used the fact that $Z_k$ and $-Z_k$ have the same
distribution.
\end{proof}
Therefore, if $\theta \geq 0$ then from Theorem \ref{thm:bax} we have \begin{equation} \label{eqn:toy_bound} ||\mathcal{L}(X_n|X_0, \theta) - \pi||_V \leq M \gamma^n V(X_0) = M \gamma^n
(1+X_0^2),\end{equation}
with $M$ and $\gamma$ computed for reversible and positive
Markov chains (see Appendix \ref{sec: Bax form rev pos} for
formulas). For $\theta < 0$ we get the same bound (\ref{eqn:toy_bound}) with exactly the same $M, \gamma$ by Lemma \ref{lemma: contracting normals
convergence} and the fact that $V(x)$ is symmetric.

The choice of $V(x) = 1+ x^2$ allows for
confidence estimation of $\int_{\stany}f(x)
\pi(dx)$ if $|f^2|_V < \infty$ for the possibly unbounded function
$f.$ In particular the MCMC works for all linear functions on
$\stany.$ We take $f(x) = x$ where $|f^2|_V = 1$ as an example. We
have to provide parameters and constants required for Theorem
\ref{thm:eps-alpha}. In this case the optimal starting point is
$X_0=0$ since it minimizes $V(x).$ Although in this example we can compute $\pi V = 2$ and $|f_c^2|_V = 1,$ we also consider bounding $\pi V$ and  $|f_c^2|_V$ using
Lemma \ref{lemma: pi V bound} and Lemma \ref{lemma: f_c^p V-norm} respectively.

\begin{small}\begin{center}
\begin{tabular}[!htb] {|c@{  }|@{  }c@{  }|@{  }c@{   }c@{   }c@{   }c@{  }|@{  }c@{   }c@{   }c@{   }c@{  }|@{  }c@{   }c@{   }c@{   }c|} \hline
& & \multicolumn{4}{c}{setting 1} & \multicolumn{4}{|c|}{setting 2} & \multicolumn{4}{c|}{reality} \\
$\alpha$ & algorithm & $m$ & $t$ & $n$ & total cost & $m$ & $t$ & $n$ & total cost & $m$ & $t$ & $n$ & total cost
\\ \hline
$.1$ & one walk &  1 & 218 & 6.46e+09 & 6.46e+09 & 1 & 229 & 1.01e+08& 1.01e+08 & 1 & 0 & 811 & 811 \\
 & MA &  - & - & - & - & - &- & -& - & - & - & - & - \\ \hline
$10^{-3}$ & one walk &  1 & 218 & 6.46e+11& 6.46e+11 &  1 & 229 & 1.01e+10& 1.01e+10 & 1 & 0 & 3248 & 3248\\
& MA & 15 & 218 & 5.40e+09& 8.10e+10 &  15 & 229 & 8.45e+07& 1.27e+09 & 7 & 0 & 726 & 5082 \\ \hline
$10^{-5}$ & one walk &  1 & 218 & 6.46e+13& 6.46e+13 &  1 & 229 & 1.01e+12& 1.01e+12 & 1 & 0 & 5853 & 5853 \\ 
& MA &  27& 218 & 5.40e+09& 1.46e+11 & 27& 229 & 8.45e+07& 2.28e+09& 13 & 0 & 726 & 9438 \\ \hline
\end{tabular}
\\
\smallskip
\begin{tabular}{p{12cm}}Table 1. Bounds for the \emph{one walk algorithm} and the \emph{median of averages Algorithm \ref{Alg: median}} (MA) for $\theta = .5,$ precision parameter $\varepsilon = .1$ and different values of the confidence parameter $\alpha.$ Baxendale's $V-$uniform ergodicity parameters in this example are $\rho = .895,\; \rho_2 = .899.$ Optimizing the total simulation cost results in $\gamma = .915,\; \gamma_2 = .971, \; M=3.64e+04,\; M_2 = 748.$ \textbf{Setting 1} uses \mbox{Lemmas \ref{lemma: pi V bound}} \mbox{and \ref{lemma: f_c^p V-norm},} whereas in \textbf{setting 2}, $\pi V$ and $|f_c^2|_V$ are computed directly. The bounds are compared to \textbf{reality} obtained empirically in a simulation study.\end{tabular}\end{center}\end{small}

Examples of bounds for $t$ and $n$ for the one walk estimator, or
$t, $ $n$ and $m$ for the median of averages (MA) estimator are
given in Table 1. The bounds
are computed for $C=[-d,d]$ with $d=1.6226$ which minimizes $\rho_2$ (rather than
$\rho$) for $\theta = 0.5.$ Then a grid search is performed to
find optimal values of $\gamma$ and $\gamma_2$ that minimize the
total simulation cost. Note that in Baxendale's result, the constant $M$
depends on $\gamma$ and goes relatively quickly to $\infty$ as
$\gamma \to \rho.$ This is the reason why optimal $\gamma$ and
$\gamma_2$ are far from $\rho$ and $\rho_2$ and turns out to
be the bottleneck of Baxendale's bounds in applications (c.f. Remark \ref{rem_leading_term}). Also for small $\alpha
= 10^{-5},$ the $m=27$ shorter runs have
a significantly lower bound on the required total simulation effort then the single long run. MA is thus more mathematically tractable. However, in reality MA is about $\pi/2$ times less efficient then the one walk estimator - a phenomenon that can be inferred from the standard asymptotic theory. 
%
%
%

R functions for computing this example and also the general bounds
resulting from Theorem \ref{thm:eps-alpha} are available at
http://www2.warwick.ac.uk/fac/sci/statistics/staff/research/latuszynski/

\section{Concluding Remarks}

The main message of our paper is a very positive one: current
theoretical knowledge of Markov chains reached the stage when for
many MCMC algorithms of practical relevance applied to difficult
problems, i.e. estimating expectations of unbounded functions, we
are able to provide a rigorous, nonasymptotic, a priori analysis
of the quality of estimation. This is much more then the often
used in practice visual assessment of convergence by looking at a
graph, more sophisticated a posteriori convergence diagnostics,
bounding only burn in time or even using asymptotic confidence
intervals, and should replace it, where possible.

The bounds derived in our paper are admittedly conservative, as observed in \mbox{Section \ref{sec: Toy Example}}. We note that this is the case also for explicit bounds on convergence in total variation norm established under drift conditions. Nevertheless drift conditions remain the main and most universal tool in obtaining nonasymptotic results for general state space Markov chains.

For regenerative algorithms alternative bounds established in \cite{LaMiaNie} are typically tighter then those resulting from our Section \ref{sec:eps-alpha}. However, the algorithms proposed there are more difficult to implement in practically relevant examples. 

\appendix
\section{Formulas for $\rho$ and M} \label{sec:Bax form}
For the convenience of the reader we repeat here the formulas from \cite{Bax} that play a key role in our considerations. 


In the sequel the term \textit{atomic case} and \textit{nonatomic
case} refers to $\tilde{\beta}=1$ and $\tilde{\beta} < 1$
respectively. If $\tilde{\beta} < 1,$ define
\[
\alpha_1 = 1+ \frac{\log\frac{K-\tilde{\beta}}{1-\beta}}{\log
\lambda^{-1}}, \quad \alpha_2 = \left\{\begin{array}{ll} 1, &
\textrm{if } \nu(C)=1, \\
1 + \frac{\log\tilde{K}}{\log \lambda^{-1}}, & \textrm{if }
\nu(C)+\int_{C^c} V d \nu \leq \tilde{K},
\\
1 + \big(\log\frac{K}{\tilde{\beta}}\big) \big/
(\log\lambda^{-1}), & \textrm{otherwise.}
\end{array} \right.
\]
Then let
\[
R_0 = \min\{\lambda^{-1}, (1-\tilde{\beta})^{-1/\alpha_1}\},
\qquad L(R) = \left\{ \begin{array}{lll}
\frac{\tilde{\beta}R^{\alpha_2}}{1-(1-\tilde{\beta})R^{\alpha_1}},
& \textrm{if} & 1 < R < R_0,\\
\infty & \textrm{if} & R=R_0. \end{array} \right.
\]
\subsection{Formulas for general operators} \label{sec: Bax form
general}
 For $\beta > 0,$ $ R> 1$ and $L>1,$ let $R_1=R_1(\beta,
R, L)$ be the unique solution $r \in (1, R)$ of the equation
\[
\frac{r-1}{r(\log(R/r))^2}=\frac{e^2\beta (R-1)}{8(L-1)}
\]
and for $1 < r < R_1,$ define
\[
K_1(r,\beta, R, L) = \frac{2 \beta + 2 (\log N)(\log (R/r))^{-1} -
8 N e^{-2} (r-1)r^{-1}(\log (R/r))^{-2}}{ (r-1) [\beta - 8N e^{-2}
(r-1) r^{-1} (\log (R/r))^{-2}]},
\]
where $N = (L-1)/(R-1).$

For the \textit{atomic case} we have $\rho = 1/R_1(\beta,
\lambda^{-1},\lambda^{-1} K)$ and for $\rho < \gamma < 1,$
\begin{eqnarray}
M & = & \frac{\max(\lambda, K-\lambda/\gamma)}{\gamma-\lambda} +
\frac{K(K-\lambda/\gamma)}{\gamma (\gamma - \lambda)}
K_1(\gamma^{-1}, \beta, \lambda^{-1}, \lambda^{-1}K) \nonumber \\
\label{eqn: M general atomic} && + \frac{(K-\lambda / \gamma)
\max(\lambda, K-\lambda)}{(\gamma - \lambda)(1-\lambda)} +
\frac{\lambda(K-1)}{(\gamma - \lambda)(1-\lambda)}.
\end{eqnarray}
For the \textit{nonatomic case} let $\tilde{R}=
\textrm{arg}\max_{1<R<R_0} R_1(\beta, R, L(R)).$ Then we have
$\rho = 1/R_1(\beta, \tilde{R},L(\tilde{R}))$ and for $\rho <
\gamma < 1,$
\begin{eqnarray}
M & = & \frac{\gamma^{-\alpha_2 -1}(K\gamma-\lambda)}{(\gamma -
\lambda)[1-(1-\tilde{\beta})\gamma^{-\alpha_1}]^2} \times \left(
\frac{\tilde{\beta} \max(\lambda, K-\lambda)}{1-\lambda} +
\frac{(1-\tilde{\beta})(\gamma^{-\alpha_1}-1)}{\gamma^{-1} - 1}
\right)
 \nonumber \\
\nonumber
 && +
\frac{\max(\lambda, K-\lambda/\gamma)}{\gamma-\lambda} +
\frac{\tilde{\beta}\gamma^{-\alpha_2
-2}K(K\gamma-\lambda)}{(\gamma -
\lambda)[1-(1-\tilde{\beta})\gamma^{-\alpha_1}]^2}
K_1(\gamma^{-1}, \beta, \tilde{R}, L(\tilde{R}))
\\ \nonumber
&&+ \frac{\gamma^{-\alpha_2}\lambda
(K-1)}{(1-\lambda)(\gamma-\lambda)[1-(1-\tilde{\beta})\gamma^{-\alpha_1}]}
+\frac{K[K\gamma-\lambda-\tilde{\beta}(\gamma-\lambda)]}{\gamma^2
(\gamma - \lambda)[1-(1-\tilde{\beta})\gamma^{-\alpha_1}]}
 \\ \label{eqn: M general nonatomic}
  && + \frac{K-\lambda
-\tilde{\beta}(1-\lambda)}{(1-\lambda)(1-\gamma)}
 \left((\gamma^{-\alpha_2}-1) +
 (1-\tilde{\beta})(\gamma^{-\alpha_1}-1)/\tilde{\beta}
 \right).
\end{eqnarray}

\subsection{Formulas for self-adjoint operators} \label{sec: Bax form
reversible} A Markov chain is said to be reversible with respect
to $\pi$ if
$\int_{\stany} Pf(x) g(x) \pi(dx) = \int_{\stany} f(x) Pg(x) \pi
(dx) $ for all $f, g \in L^2(\pi).$ For reversible Markov chains
the following tighter bounds are available.

For the \textit{atomic case} define
\[
R_2 = \left\{\begin{array}{lll} \min \left\{ \lambda^{-1}, r_s
\right\},  & \textrm{if} & K
> \lambda + 2\beta, \\ \lambda^{-1}, & \textrm{if} & K \leq \lambda +
2\beta,
\end{array} \right.
\]
where $r_s$ is the unique solution of $1 + 2 \beta r = r^{1+(\log
K)(\log \lambda^{-1})}.$
Then $\rho = R_2^{-1}$ and for $\rho < \gamma < 1$ take $M$ as in
(\ref{eqn: M general atomic}) with $K_1(\gamma^{-1}, \beta,
\lambda^{-1}, \lambda^{-1}K)$ replaced by $K_2 = 1+
1/(\gamma-\rho).$

For the \textit{nonatomic case} let
\[
R_2 = \left\{\begin{array}{lll} r_s, & \textrm{if} & L(R_0) > 1 +
2\beta R_0,
\\R_0, & \textrm{if} & L(R_0) \leq 1 + 2\beta R_0,
\end{array} \right.
\]
where $r_s$ is the unique solution of $1 + 2 \beta r = L(r).$
Then $\rho = R_2^{-1}$ and for $\rho < \gamma < 1$ take $M$ as in
(\ref{eqn: M general nonatomic}) with $K_1(\gamma^{-1}, \beta,
\tilde{R}, L(\tilde{R}))$ replaced by $K_2 = 1 +
\sqrt{\tilde{\beta}}/(\gamma-\rho).$

\subsection{Formulas for self-adjoint positive operators}
\label{sec: Bax form rev pos} A Markov chain is said to be
positive if $\int_{\stany}Pf(x)f(x)\pi(dx) \geq 0$ for every $f
\in L^2(\pi).$ For reversible and positive markov chains take
$M$'s as in Section \ref{sec: Bax form reversible} with $\rho =
\lambda$ in the \textit{atomic case} and $\rho = R_0^{-1}$ in the
\textit{nonatomic case.}
\end{document}